\documentstyle[twocolumn,epsf]{jpsj}

\catcode`\@=11

\def\simle{\mathrel{\mathpalette\@versim<}}   
\def\simge{\mathrel{\mathpalette\@versim>}}   
\def\@versim#1#2{\lower2.5pt\vbox{\baselineskip0pt \lineskip-.5pt
   \ialign{$\m@th#1\hfil##\hfil$\crcr#2\crcr\sim\crcr}}}

\newcommand{\bequ}{ \begin{equation} }
\newcommand{\eequ}{ \end{equation} }
\newcommand{\barr}{ \begin{array} }
\newcommand{\earr}{ \end{array} }
\newcommand{\beqarr}{ \begin{eqnarray} }
\newcommand{\eeqarr}{ \end{eqnarray} }

\newcommand{\baralpha}{ \begin{eqnal} \beqarr}
\newcommand{\earalpha}{ \eeqarr \end{eqnal}}

\catcode`\@=12

\def\runtitle{
}
\def\runauthor{Yukitoshi {\sc Motome} and Nobuo {\sc Furukawa}}

\def\sf{\rm}

\title{
Critical Phenomena of Ferromagnetic Transition in\\
Double-Exchange Systems
}

\author{
Yukitoshi {\sc Motome} and Nobuo {\sc Furukawa}$^{1}$ 
}
\inst{
Institute of Materials Science, University of Tsukuba,
Tsukuba, Ibaraki 305-0006 \\
$^{1}$Department of Physics, Aoyama Gakuin University, 
Setagaya, Tokyo 157-8572
}
\recdate{
}

\abst{
Critical phenomena of ferromagnetic transition at finite temperatures
are studied in double-exchange systems.
In order to investigate strong interplay
between charge and spin degrees of freedom,
Monte Carlo technique is applied to include
fluctuations in a controlled and unbiased manner.
By using finite-size scaling analysis,
critical exponents and transition temperature
are estimated for a model with Ising spin symmetry in two dimensions.
The obtained exponents are far distinct from the mean-field values,
but consistent with those of spin models with short-range exchange interactions.
The universality class of this transition belongs to
that of short-range interaction with the same spin symmetry.
We also discuss the case for three dimensions.
The results are compared with experimental results
in perovskite manganites which show colossal magnetoresistance.
}

\kword{
double-exchange system, ferromagnetic transition, critical exponents,
finite-size scaling, Monte Carlo, polynomial expansion,
colossal-magnetoresistance manganites
}

\begin{document}
\sloppy
\maketitle


Perovskite manganites which show the colossal magnetoresistance
have revived interests in the double-exchange (DE) mechanism.
\cite{Zener1951}
Recent intensive experiments have improved the precision of data remarkably,
\cite{Ramirez1997}
which now makes an issue of detailed comparisons between experiment and theory
in a quantitative way.
Although theoretical survey on the DE systems has a long history,
many aspects at finite temperatures have not been fully understood thus far.
\cite{Furukawa1999}
For instance, the critical temperature of the metallic-ferromagnetic transition
has been controversial until quite recently.
\cite{Motome2000}

What makes it difficult to describe the system
is strong spin fluctuation.
The system at issue is a strongly-correlated electron system
since Hund's-rule coupling between conduction electrons and localized spins
is much larger than the kinetic bandwidth.
Electrons move around and align localized spins in a parallel way
to gain the kinetic energy, which is an origin of the DE interaction.
\cite{Zener1951}
At finite temperatures, especially near the transition,
spin fluctuations strongly modify
the kinetics of electrons and vice versa
through the strong interplay between charge and spin degrees of freedom.

One of the most fundamental and challenging problems is critical phenomenon
of the ferromagnetic transition in this system.
The DE interaction has a distinguishable property
compared to the ordinary exchange interactions in localized-spin models
such as the Heisenberg model.
Since the interaction is originated from the motion of conduction electrons,
the effective range of the interaction may depend on 
kinetics of electrons and hence on temperatures crucially
through the strong fluctuations.
It is not clear how the interaction is renormalized near the critical point
and what universality class governs the magnetic transition.

\begin{figure}
\epsfxsize=6cm
\centerline{\epsfbox{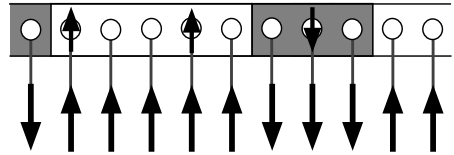}}
\caption{
Schematic picture of the double-exchange model
in the limit of strong Hund's-rule coupling.
White and gray areas describe ferromagnetic domains where electrons are
localized within.
}
\label{fig:DE}
\end{figure}

To be more specific, we consider a situation
where spin configuration forms ferromagnetic domains
as illustrated in Fig.~\ref{fig:DE}, as an example.
In the limit of strong Hund's-rule coupling,
conduction electrons are completely parallel to localized spins
and confined within the domains.
In this system, the total energy 
is given by a sum of energy eigenvalues $\varepsilon_\nu $,
$E_{\rm tot}= \sum \varepsilon_\nu $, where 
summation is restricted to occupied states.
Energy eigenvalues $\varepsilon_\nu$ for electrons
confined within domains  strongly depend on
the  sizes as well as the shapes of the domains.
In order to describe $E_{\rm tot}$ as a function of
spin configuration, it is necessary to introduce
long-range two-spin interactions as well as multiple-spin interactions
which depend  on sizes and shapes of  ferromagnetic clusters.
This is a marked contrast to the ordinary spin models
with short-range exchange interactions,
where the total energy depends only 
on surface volumes of domain boundaries.

\vspace*{-5mm}
The size of domains grows up and the shape fluctuates strongly
as the system approaches to the critical point.
In the DE systems, it is difficult to describe
the renormalization of the effective magnetic interaction near the transition,
whether it is renormalized to a short-range interaction, 
or the long-ranged and multiple-spin interaction
become relevant to cause mean-field-like transition
through suppression of fluctuations.
Thus, it is highly nontrivial and interesting
to determine the universality class of the ferromagnetic transition 
of the DE origin.

Recently, we studied the critical phenomena
in the three-dimensional (3D) DE model.
\cite{Motome2000}
The critical exponents are examined
by using the finite-size scaling analysis on the Monte Carlo (MC) data
which is in principle an  unbiased method.
Our data scale more consistently to the short-range Heisenberg exponents
than to those of mean-field results.
However,  the data are not precise enough
to determine the universality class in detail,
mainly because the system size
is still not large enough due to the limitation
in computational time.

In this letter, we examine critical phenomena of
the metallic ferromagnetism  in the DE systems,
through a precise estimation  for critical exponents.
For this purpose,
we study a simplified DE model 
which still captures the essence of the DE mechanism.
By analyzing the MC data using the finite-size scaling,
we estimate the critical exponents as well as the
critical temperature.
We compare the results with those for
spin models with short-range type interactions as well as
mean-field values.


In order to perform calculations on a larger scale more precisely,
we study minimum models which include the DE mechanism and
show the ferromagnetic transition at finite temperatures.
One of such models is the DE model
where electrons are coupled to spins with Ising symmetry in two dimensions.
The Hamiltonian is given by
\begin{equation}
\label{eq:H_org}
{\cal H} = -t \sum_{<ij>, \sigma}
c^{\dagger}_{i \sigma} c_{j \sigma}
- J_{H} \sum_{i} \sigma^{z}_{i} \cdot S_{i},
\end{equation}
where $c_{i \sigma}$ ($c_{i \sigma}^{\dagger}$) annihilates (creates)
a $\sigma$-spin electron at site $i$,
$J_{\rm H}$ is the Hund's-rule coupling,
$\sigma^{z}$ is the $z$-component of the spins of conduction electrons, and
$S$ describes the localized Ising spin which takes $S=\pm 1$.
The summation in the first term is taken for nearest-neighbor sites.
For simplicity, we consider the limit of strong Hund's-rule coupling,
$J_{H} \rightarrow \infty$.
Then the electron hopping between sites $i$ and $j$
is allowed only when the localized spins are parallel, that is, $S_{i} = S_{j}$.
The model becomes equivalent to
a spinless-fermion model with hopping integrals
which depend on the spin configurations:
\begin{equation}
\label{eq:H}
{\cal H} = -  \sum_{<ij>} \frac{t}{2}(1 + S_{i}S_{j}) \ 
\tilde{c}^{\dagger}_{i} \tilde{c}_{j},
\end{equation}
where $\tilde{c}_{i}$ and $\tilde{c}^{\dagger}_{i}$
are the spinless-fermion operators.
Note that model (\ref{eq:H}) is simplified compared to the original model.
Nevertheless, it still includes the essence of the DE mechanism;
the kinetics of electrons is strongly correlated with spin degrees of freedom and
energetically favors ferromagnetism.

We study model (\ref{eq:H}) by an improved MC method.
\cite{Motome1999}
The most time-consuming part in the standard MC algorithm is
the exact diagonalization of the fermion Hamiltonian for each spin configuration.
\cite{Yunoki1998}
In the present algorithm, it is replaced by
moment expansion of density of states.
This reduces the cost of computational time considerably,
while the truncation error falls off exponentially.
The algorithm has another advantage to reduce computational time;
the moment expansion is performed on parallel computers efficiently.

In the following, the ferromagnetic transition in model (\ref{eq:H})
is studied at the electron filling
$x = \langle \sum_{i} \tilde{c}^{\dagger}_{i} \tilde{c}_{i} \rangle = 0.5$.
Here, the bracket denotes the thermal average for the grand canonical ensemble.
A closed-shell condition is chosen to obtain rapid convergence of MC samplings and
systematic extrapolation of system sizes.
\cite{Motome1999}
We study the system with square lattice up to $20 \times 20$
under the periodic boundary condition in one direction and
the antiperiodic boundary condition in the other direction.
We take the half-bandwidth $W$ of noninteracting electrons as an energy unit;
$W = 4t$ in two dimensions.
The moment expansion of density of states is taken up to 20th order.
We confirm that the truncation error is negligible small in the region
$T/W \ge 0.05$ in model (\ref{eq:H})
in comparison with the results by exact diagonalization method.
We have typically run 100,000 MC samplings for measurements
after 10,000 MC steps for thermalization.
Measurement is divided into five bins
to estimate the statistical error by the variance among the bins.

\begin{figure}
\epsfxsize=7cm
\centerline{\epsfbox{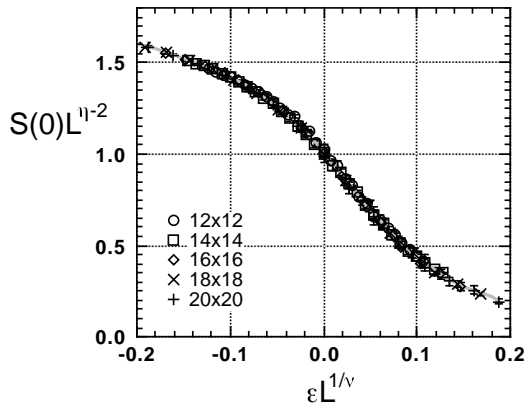}}
\caption{
The best-fit result of the scaling plot for the Monte Carlo data.
The gray curve in the figure is the polynomial fit.
}
\label{fig:scaling}
\end{figure}

In order to determine the critical exponents and
the critical temperature $T_{\rm c}$,
we use the finite-size scaling analysis on the ferromagnetic component of
the spin structure factor.
The spin structure factor is defined by
\begin{equation}
\label{eq:S(k)}
S(\mib{k}) = \frac{1}{N} \sum_{ij} \langle S_{i} S_{j} \rangle
e^{{\rm i} \mib{k} \cdot \mib{r}_{ij}},
\end{equation}
where $N$ is the total number of sites.
If we assume the hyperscaling hypothesis,
the ferromagnetic component $S(\mib{k}=0)$ satisfies
the scaling relation near the ferromagnetic transition as
\begin{equation}
\label{eq:S(0)scaling}
S(0) L^{\eta-2}  = f ( \epsilon L^{1/\nu} ),
\end{equation}
where $\epsilon = (T-T_{\rm c})/T_{\rm c}$,
$L$ is a linear dimension of the system and
$f$ is the scaling function.
Here $\eta$ and $\nu$ are the critical exponents,
by which we can determine the universality class
under the hyperscaling hypothesis.
For given exponents and $T_{\rm c}$,
we plot $S(0) L^{\eta-2}$ as a function of $\epsilon L^{1/\nu}$ and
fit the data by a polynomial function $f_{\rm fit}$
which approximately gives the scaling function $f$.
The best estimate for exponents and $T_{\rm c}$ is given
by minimizing the systematic deviations of the data from the scaling function.
We define the deviation function
by differences between the fit and the MC data as
\begin{equation}
\label{eq:R}
R(\eta, \nu, T_{\rm c}) = \sum_{n}
\left[ \frac{y_{n} - f_{\rm fit}(x_{n})}{f_{\rm fit}(x_{n})} \right]^{2},
\end{equation}
where $y_{n} = S_{n}(0) L_{n}^{\eta-2}$ and
$x_{n} = \epsilon_{n} L_{n}^{1/\nu}$.
Here the subscript $n$ labels MC data for different parameters $T$ and $L$.
We apply the scaling of eq.~(\ref{eq:S(0)scaling})
on MC data for the systems with $L = 12, 14, 16, 18$ and $20$
for $0.05 \le T/W \le 0.065$.
We use the fitting function $f_{\rm fit}$ of
the polynomials up to 12th order.
For convenience of later discussions,
hereafter we use the exponent $\beta$
which is related with $\eta$ and $\nu$ by $\beta = \eta\nu/2$
in two dimensions.
The best estimate which minimizes $R$ is obtained at
$\beta = 0.09$,
$\nu = 0.9$ and
$T_{\rm c}/W = 0.058$.
The scaling behavior at these values is shown in Fig.~\ref{fig:scaling}.
All the MC data are scaled well to a single universal function
within the errors.

\vspace*{-5mm}

\begin{figure}
\epsfxsize=8cm
\centerline{\epsfbox{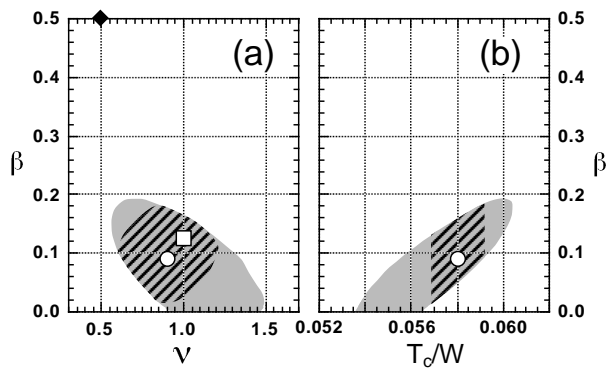}}
\caption{
Estimated values of (a) the critical exponents $\nu$ and $\beta$;
and (b) the critical temperature $T_{\rm c}$ and $\beta$.
The circle, square and diamond in the figures
denote the best-fit result, the Ising exponents and
the mean-field exponents, respectively.
The gray areas, both with and without hatches, exhibit the estimates
by the scaling analysis of the spin structure factor.
The hatched areas are the results
by using the estimate for $T_{\rm c}$ 
by the Binder parameter in addition.
See text for details.
}
\label{fig:exponents}
\end{figure}

Errors of the critical exponents are estimated in the following way.
Since the MC data have statistical errors,
the deviation function (\ref{eq:R}) also has an error $\delta R$.
In the parameter space of $(\beta, \nu, T_{\rm c})$,
there exists a region where the values $R$ are not distinct
from its minimum value within the statistical errors,
by which we define the region for the estimate of $(\beta, \nu, T_{\rm c})$.
Namely, we determine the error region by the condition
\begin{equation}
\label{eq:condR}
R - \delta R < R^{*} + \delta R^{*},
\end{equation}
where $R^{*}$ and $\delta R^{*}$ are the minimum value of $R$ and
its error at the best-fit point, respectively.
The gray regions in Fig.~\ref{fig:exponents},
including hatched areas, show the results
by making projections of the error region
onto $\beta$-$\nu$ and $\beta$-$T_{\rm c}$ planes.

\vspace*{-5mm}

As an alternative method to estimate $T_{\rm c}$ independently,
we perform the Binder parameter analysis.
\cite{Binder1981}
The Binder parameter is defined by
$g = 1 - \langle q^{4} \rangle / 3 \langle q^{2} \rangle^{2}$
where the order parameter $q$ is taken as
$\sum_{i} S_{i}$ in this case.
For $T > T_{\rm c}$ ($T < T_{\rm c}$),
the Binder parameter decreases (increases) as the system size $N$ increases.
Thus $T_{\rm c}$ is determined by a crossing point of $g$ for various $N$.

\begin{figure}
\epsfxsize=6cm
\centerline{\epsfbox{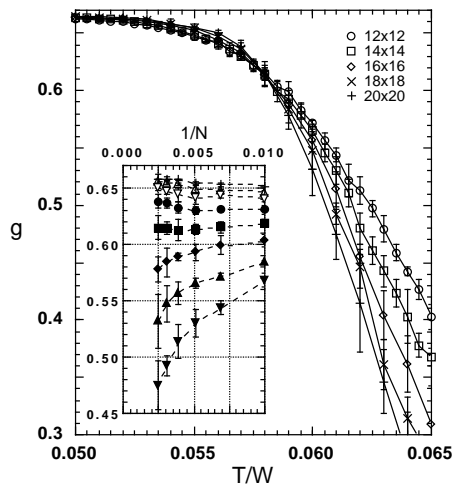}}
\caption{
Temperature dependence of the Binder parameter.
The inset is system-size dependence;
the data are for $T/W = 0.061, 0.06, 0.059, 0.058, 0.057, 0.056, 0.055$ and
$0.054$ from bottom to top.
The curves are guides to the eye.
}
\label{fig:Binder}
\end{figure}

Figure \ref{fig:Binder} shows the temperature dependence of $g$
for different system sizes.
The inset displays the system-size dependence.
The Binder parameter seems to cross at $T/W \simeq 0.058$.
The data at $T/W = 0.058$ are almost independent of $N$ as shown in the inset.
We estimate $T_{\rm c}/W = 0.058 \pm 0.001$.
This is consistent with the results in Fig.~\ref{fig:exponents}.
Using this value of $T_{\rm c}$ as an additional criterion
in the analysis of eq.~(\ref{eq:condR}),
we obtain the estimates for the critical exponents
as shown by the hatched regions in Fig.~\ref{fig:exponents}.
As a conclusion from combined analysis of finite-size scaling and Binder plots,
we obtain $\beta = 0.09 \pm 0.08$, $\nu = 0.9 \pm 0.3$ and
$T_{\rm c} = 0.058 \pm 0.001$.
In Table \ref{table:exponents}, we show a summary of the exponents.

\vspace*{-10mm}

\begin{table}
\caption{Comparison of the critical exponents.}
\begin{tabular}{@{\hspace{\tabcolsep}\extracolsep{\fill}}ccc}
 & $\beta$ & $\nu$ \\
\hline
present model & $0.09 \pm 0.08$ & $0.9 \pm 0.3$ \\
2D Ising model & $0.125$ & $1$ \\
mean-field & $0.5$ & $0.5$
\end{tabular}
\label{table:exponents}
\end{table}

\vspace*{-4mm}

\begin{figure}
\epsfxsize=6cm
\centerline{\epsfbox{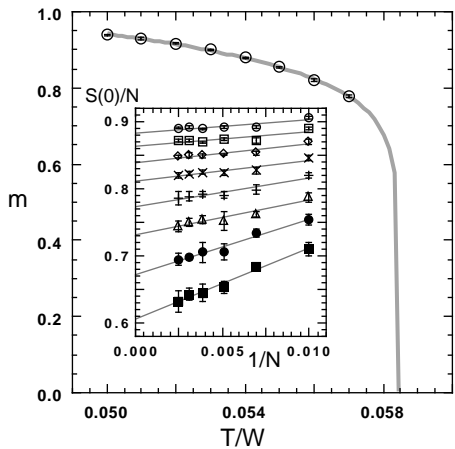}}
\caption{
Temperature dependence of the magnetization.
The curve is the least-square fit to $m \propto (T-T_{\rm c})^\beta$.
The inset shows system-size extrapolation of the spin-structure factor.
The lines are the least-square fits to $1/N$.
The data are for $T/W = 0.05, 0.051, 0.052, 0.053, 0.054, 0.055, 0.056$ and $0.057$
from top to bottom.
}
\label{fig:mag}
\end{figure}

The exponents and $T_{\rm c}$ in Fig.~\ref{fig:exponents} are crosschecked
by the scaling behavior of the magnetization.
The magnetization is calculated by
$m = [ S(\mib{k}=0) / N ]^{1/2}$.
The temperature dependence of the magnetization is shown in Fig.~\ref{fig:mag}.
The values in the thermodynamic limit are obtained
by system-size extrapolation of the spin structure factor,
which is shown in the inset.
MC data are scaled well to $1/N$, which is expected
in the ordered phase in the limit of large $N$.
We fit the magnetization by assuming the scaling relation
$m \propto (T_{\rm c} - T)^{\beta}$
and obtain $T_{\rm c}/W = 0.0584 \pm 0.0005$ and
the exponent $\beta = 0.11 \pm 0.01$.
If we assume that these data are within the critical region,
these values of $T_{\rm c}$ and $\beta$
are all consistent with the results in Fig.~\ref{fig:exponents}.
Note that in this transition it is difficult to determine $T_{\rm c}$ and $\beta$
by the magnetization data alone because of the small value of $\beta$.


Now, we discuss the results in Fig.~\ref{fig:exponents}.
For comparison,
we plot the critical exponents of the Ising model
with nearest-neighbor interaction in two dimensions;
$\beta = 0.125$ and $\nu = 1$.
\cite{Kaufmann1944, Stephenson1964}
These values are included in our estimated regions and
very close to the best-fit result.
On the contrary, the critical exponents of the mean-field approximation,
$\beta = \nu=1/2$, are far outside of the area
as shown in the figure.
These results indicate that
the universality class of 
the ferromagnetic transition in our simplified model (\ref{eq:H}) is
the same as that in models which have short-range two-spin interactions
but clearly distinct from that of the mean-field approximation.

\vspace*{-5mm}

Let us now consider the 3D DE model.
The nature of the magnetic interaction which is 
mediated by itinerant electrons are common.
Estimates for the critical exponents in the 3D DE model are similar to
the short-range Heisenberg type rather than 
the mean-field one.\cite{Motome2000}
Comparison between the present result and those for 3D DE model
strongly suggests the following:
The long-range and multiple-spin parts of the magnetic 
interactions in the  DE models
seem to be renormalized to be irrelevant.
The universality class of the ferromagnetic transition
in the DE systems should belong to that of the short-range interaction
with the same spin symmetry.

Concerning experiments in manganites, there have been many studies
to estimate the critical exponents of the ferromagnetism.
Up to now, however, 
the experimental estimates for exponents are still controversial,
including those for short-range Heisenberg interaction,
\cite{Heffner1996, Martin1996, Lofland1997a, Vasiliu-Doloc1998, Ramos1998}
the mean-field values,
\cite{Lofland1997b, Mohan1998, Schwartz2000}
and those which cannot be classified into any universality class ever known.
\cite{Ghosh1998}

From our numerical results, it is suggested that the exponents for
the short-range Heisenberg universality class should be obtained
when the transition in real materials is ascribed to the DE mechanism
as a major origin.
Further experimental studies are desired to
clarify the critical exponents of the real materials, and examine
whether the transition follows the single-parameter scaling
in the universality class of the short-range type.
These include sample-quality refinements, especially
in homogeneities of $T_{\rm c}$, as well as
appropriate treatments for the fitting within critical regions.
The experimental results should cast a crucial test
whether the DE mechanism alone can explain experimental results or not.


In summary, we have studied critical phenomena of the ferromagnetic transition
in the simplified double-exchange system numerically.
The model which has Ising spin symmetry
has been studied in the limit of strong coupling on square lattices.
We have applied an improved Monte Carlo algorithm
which reduces computational time considerably.
Critical exponents and critical temperature are estimated
by the finite-size scaling analysis on the Monte Carlo data.
The critical exponents $\beta = 0.09 \pm 0.08$ and
$\nu = 0.9 \pm 0.3$ are consistent with
those of models with short-range interactions,
but distinct from those of the mean-field approximation.
The universality class of this transition belongs to
the short-range type with the same spin symmetry in the model.


The authors thank H. Nakata for helpful support
in developing parallel-processing systems.
The computations have been performed mainly 
using the facilities in the AOYAMA+ project
(http://www.phys.aoyama.ac.jp/\\ \~{}aoyama+)
and in the Supercomputer Center, Institute for Solid State Physics,
University of Tokyo.
This work is supported by  ``a Grant-in-Aid from
the Ministry of Education, Culture, Sports, Science and Technology''.


\end{document}